# Thermodynamic reflection of particles by Schwarzschild black holes

Scott Funkhouser
National Oceanic and Atmospheric Administration,
2234 South Hobson Avenue
Charleston, SC 29405-2413

ABSTRACT

The change in the entropy caused by the quasi-static absorption of a particle of energy $\varepsilon$ by a Schwarzschild black hole (ScBH) is approximately $\varepsilon/T - s$, where $T$ is the Hawking temperature of the black hole and $s$ is the entropy of the particle. A violation of the generalized Second Law of Thermodynamics would occur if $\varepsilon/T - s < 0$, and it is plausible that particles approaching the event horizon of a ScBH may be reflected thermodynamically in some instances. The reflection probability is obtained from the standard relationship between the number of microscopic complexions and entropy. If $(\varepsilon/T) \gg 0$ and if $s$ is negligible then the new probability function is consistent with an independent expression, following from a detailed treatment of quantum particles in a Schwarzschild metric, giving the probability for an event horizon to reflect an incident particle. The manifestation of wave-like behaviors in the new probability function intimates perhaps a fundamental physical unity.

## I. *The thermodynamics of particle absorption*

Consider an isolated system $X$, subject only to internal forces, that consists of a Schwarzschild black hole (ScBH) and a particle that is approaching the black hole and that is, classically, bound to the black hole by gravitation. Let the boundary of the system be an arbitrary sphere that is centered on the black hole and whose radius is very large in comparison to the event horizon of the black hole, so that the system is essentially closed. If the system behaves according to classical general relativity and classical particle physics then the particle would be drawn to the black hole, never to escape, and would reach the event horizon only after an infinite time. In modern, quantum-mechanical models of the interactions between particles and black holes there exist non-classical behaviors that could cause the system to evolve in a manner that differs significantly from the classical scenario.

The non-classical physics that could be important in a system like $X$ may be categorized broadly as being associated with either the particle or the black hole. The purpose of this work is to investigate a basic implication of the generalized Second Law of thermodynamics (GSL), which is a non-classical aspect of black holes, for the absorption of a particle characterized only by an entropy and energy. The present investigation reveals a surprising unity between the non-classical behaviors of black holes with respect to classical particles and the non-classical behaviors of quantum particles with respect to classical black holes.

In this Section is derived a critical condition, in terms of the entropy and energy of the particle and the temperature of the black hole, characterizing the regime for which absorption of a particle would violate the GSL. If the condition is not satisfied then absorption may be statistically suppressed, and it is allowed that incident particles could be reflected thermodynamically by the event horizon of a ScBH. Basic statistical principles are then employed to obtain a function giving the probability for thermodynamic reflection. Although no explicitly quantum-mechanical properties of the particle are considered in its derivation, the new probability function features a certain wave-like behavior. In Section II, thermodynamic reflection is compared to the independently proposed phenomenon of quantum mechanical reflection, which follows from the behaviors of quantum fields in the Schwarzschild metric. The respective probabilities are found to converge, and the consistency between the two phenomena is discussed in terms of a thermodynamic interpretation of the wave-like behavior of matter.

Following from the pioneering work of Bekenstein and Hawking, it is possible to specify, for a given ScBH, an effective temperature and entropy that have physical significance in the context of certain thermodynamic laws [1], [2], [3]. The articulation of the thermodynamic properties of black holes has introduced a sea change in modern physics, bearing revolutionary implications for thermodynamics and cosmology, among other fields of study [4],[5]. The entropy $S_{bh}(R)$ of a ScBH is given by

$$S_{bh}(R) = k_B \frac{\pi R^2}{l_P^2}, \tag{1}$$

where $R$ is the event horizon, $l_P$ is the Planck length and $k_B$ is the Boltzmann constant [6]. The effective temperature $T_{bh}(M)$ of a ScBH is given by

$$T_{bh}(M) = \frac{1}{k_B} \frac{m_P^2 c^2}{8\pi M}, \tag{2}$$

where $M$ is the mass of the black hole and $m_P$ is the Planck mass [7]. Suppose that a certain ScBH is contained within some very large, closed system. Let the entropy of the black hole be $S_{bh}$ at some time of reference and let $S_{out}$ represent the total entropy of the remainder of the system outside of the black hole at that time. Consider some thermodynamic transaction in which the change in the entropy of the black hole is $\Delta S_{bh}$ and the entropy of the remainder of the system changes by an amount $\Delta S_{out}$. It follows from extending the Second Law of Thermodynamics (SL) to address the entropy of black holes that

$$\Delta S_{bh} + \Delta S_{out} \geq 0. \tag{3}$$

The requirement (3) is the most basic statement of the GSL [1][6]. The conclusions of this present work follow directly from (3) and (1). It is interesting to note, however, that the temperature (2) emerges naturally in the following.

Let the energy $\varepsilon$ of the particle in a system like $X$ be much smaller than $M_1 c^2$, where $M_1$ is the initial mass of the black hole and $c$ is the vacuum-speed of light. It is presumed that the decrease in the mass of the black hole due to the emission of Hawking radiation during the interaction with the particle is negligible. It is also presumed that the interaction between the particle and the black hole does not change significantly the physical state of the black hole, and the black hole may be therefore treated as quasi-static during the evolution of the particle. Suppose that the particle is absorbed, effectively, by the black hole. The mass of the black hole would increase accordingly by $\varepsilon/c^2$, and the event horizon would become

$$R_2 = R_1 + \frac{2G\varepsilon}{c^4}, \tag{4}$$

where $R_1 = 2GM_1/c^2$ is the initial horizon. The corresponding increase $\Delta S_{bh}$ in the entropy of the black hole is necessarily $S_{bh}(R_2) - S_{bh}(R_1)$, and it follows from (1) that

$$\Delta S_{bh} = \frac{\pi k_B}{l_P^2} \left( R_2^2 - R_1^2 \right). \tag{5}$$

Since $\varepsilon \ll M_1 c^2$, the corresponding increase $\Delta R = R_2 - R_1$ in the event horizon is very small with respect to $R_1$, and (5) is therefore very well approximated by

$$\Delta S_{bh} \cong 2\pi k_B \frac{R_1 \Delta R}{l_P^2}. \tag{6}$$

It follows from (4) and (2) that $\Delta S_{bh}$ is given by

$$\Delta S_{bh} \cong 8\pi k_B \frac{M_1 \varepsilon}{m_P^2 c^2} = \frac{\varepsilon}{T_1}, \tag{7}$$

where $T_1$ is defined as $T_{bh}(M_1)$. Note that (7) gives the increase in the entropy of a ScBH resulting from the absorption of any quantity of energy $\varepsilon$, provided that $\varepsilon \ll M_1 c^2$ and that the black hole is otherwise quasi-stationary [8]. The result in (7) is noteworthy since it is analogous to the standard thermodynamic expression for the change in entropy of a system that absorbs a quantity of energy at a constant temperature.

Let $s$ represent the total entropy attributed to the particle when it is located very far from the black hole. The absorption of the particle by the black hole causes the entropy of the region of the system that is outside of the black hole to change by an amount $\Delta S_{out} \cong -s$. The net change in the entropy $\Delta S_a$ caused by the absorption of the particle by a black hole in a system like $X$ is thus

$$\Delta S_a = \frac{\varepsilon}{T_1} - s. \tag{8}$$

If $\Delta S_a$ is negative then absorption of the particle would violate the GSL. It is important to note that the Second Law of Thermodynamics (SL) and its generalization addressing black holes represent essentially probabilistic arguments. Strictly interpreted, it is not *impossible* for violations of the SL or GSL to occur, but only *improbable*. If the GSL is physically meaningful for the present scenario then absorption must be unlikely in those scenarios when $\Delta S_a < 0$. It is therefore plausible that a black hole may effectively reflect an incident particle in certain scenarios in order to preserve the GSL.

According to classical models, a particle in a system like $X$ will be effectively absorbed by the black hole, and there is consequently only one final configuration allowed to the system. If the preceding analysis is meaningful then it is appropriate to allow an additional final configuration in which the particle has not been absorbed and remains outside of the black hole. Following the standard statistical prescription, the probability $P_j$ for the system to be found in a given macrostate $\Sigma_j$ is proportional to the number $W_j$ of microstates, or "complexions", encompassed by $\Sigma_j$. In general, $W_j$ is given by

$$W_j = \exp\left(\frac{S_j}{k_B}\right), \tag{9}$$

where $S_j$ is the total entropy of $\Sigma_j$. Let $\Sigma_a$ represent the macrostate in which the black hole has effectively absorbed the particle. Let $\Sigma_{na}$ represent the macrostate in which the particle was not absorbed but is found within the system and outside of the black hole. The number $W_a$ of complexions of $\Sigma_a$ is given by

$$W_a = \exp\left(\frac{S_1}{k_B} + \frac{\varepsilon}{k_B T}\right), \tag{10}$$

and the number $W_{na}$ of complexions of $\Sigma_{na}$ is given by

$$W_{na} = \exp\left(\frac{S_1}{k_B} + \frac{s}{k_B}\right). \tag{11}$$

The probability $P_a$ for the system to be found in the macrostate $\Sigma_a$ is proportional to $W_a$, and the probability $P_{na}$ associated with $\Sigma_{na}$ is proportional to $W_{na}$. Note that the ratio $P_a/P_{na}$ must be equal to $W_a/W_{na}$, and it is therefore appropriate to define $A$ such that $P_a \equiv AW_a$, $P_{na} \equiv AW_{na}$ and $P_a+P_{na}=1$. It follows readily from the associated algebra that

$$A = \exp\left(-\frac{S_1}{k_B} - \frac{s}{k_B}\right) \cdot \left[1 + \exp\left(\frac{\varepsilon}{k_B T_1} - \frac{s}{k_B}\right)\right]^{-1}. \quad (12)$$

The probabilities $P_a$ and $P_{na}$ are therefore given by

$$P_a = \frac{\exp(\Delta S_a / k_B)}{1 + \exp(\Delta S_a / k_B)} \quad (13)$$

and

$$P_{na} = \frac{1}{1 + \exp(\Delta S_a / k_B)}, \quad (14)$$

where $\Delta S_a = (\varepsilon/T_1) - s$ is the change in entropy of the system that would occur if the particle were absorbed.

In general, the probability for the entropy of a system to change by an amount $\Delta S$ should increase with increasing $\Delta S$, and should decrease as $\Delta S$ decreases. If $\Delta S<0$ then the associated probability should vanish as the magnitude of $\Delta S$ becomes infinitely large. Likewise, if $\Delta S>0$ then the associated probability should approach unity as $\Delta S$ becomes infinitely large. As $\Delta S_a$ approaches (positive) infinity, the probability for absorption (13) approaches unity and the reflection probability (14) vanishes. Alternatively, as $\Delta S_a$ approaches negative infinity, (14) approaches unity and (13) vanishes. If $\Delta S_a=0$ then (13) and (14) are equal. The probabilities (13) and (14) are thus consistent with the basic expectations following from the SL.

*II. Discussion of physical implications*

Certain features of the analysis in Section I may intimate unanticipated physical relationships among thermodynamic principles and other natural laws. In particular, the phenomenon of thermodynamic reflection may be related inherently to an analogous quantum mechanical behavior. The probabilities (13) and (14) were obtained from conventional statistical thermodynamics by specifying only the energy and entropy of the incident particle, the initial entropy of the black hole and the change in the entropy of the black hole, $\Delta S_{bh}$, that would follow from absorbing the particle. The articulation of $\Delta S_{bh}$ followed from the definition of the black hole entropy in (1), which is non-classical. In contrast, the evolution of a system like $X$ has also been modeled by combining quantum field theory and classical general relativity. It has been established that such models allow a particle to approach the event horizon of a classical ScBH and, instead of being absorbed, return from the vicinity of the horizon in a finite time [9] – [14]. The probability for the particle to be reflected in that manner is given by

$$P_r = \exp\left(\frac{-\varepsilon}{k_B T_1}\right), \quad (15)$$

where $\varepsilon$ is the energy of the particle and $T_1$ is the initial temperature of the black hole [9]—[14]. Remarkably, the probability for thermodynamic reflection in (14) is identical to (15) in the limit as $s$ vanishes and $(\varepsilon/T_1)$ as becomes arbitrarily large.

Note that the reflection probabilities (15) and (14) both increase as $\varepsilon$ decreases with respect to $T_1$ (for a fixed $s$). Classically, decreasing the energy of the particle should facilitate absorption, with all else constant. In the context of quantum mechanics, however, there exists a physical explanation for the non-classical behaviors of (15) and (14) with respect to $\varepsilon/T_1$. It is instructive to note that $\varepsilon/T_1$ may be expressed alternatively as

$$\frac{\varepsilon}{T_1} = 4\pi k_B \frac{R_1}{(\hbar c/\varepsilon)}, \tag{16}$$

where $\hbar$ is the Planck quantum. The term $\hbar c/\varepsilon$ represents the quantum mechanical wavelength associated with the energy $\varepsilon$. If the energy of the particle is $\sim mc^2$, where $m$ is the mass of the particle, then $\hbar c/\varepsilon$ is also proportional to the Compton wavelength of the particle. Decreasing $\varepsilon/T_1$ corresponds to increasing the ratio of the characteristic wavelength of the particle to the event horizon. In general, the quantum mechanical probability for a particle to become bound within a potential well must decrease as the characteristic wavelength of the particle increases with respect to the characteristic size of the well, all else constant. If $\Delta S_a \gg 0$ and $(\varepsilon/T_1) \gg s$, which is the regime in which (14) and (15) converge, the characteristic wavelength of the particle is very small with respect to the horizon of the black hole, and the entropy of the particle is negligible. The interaction between the particle and black hole in that regime is nearly classical. Both (14) and (15) vanish in the classical limit, which is consistent with the Correspondence Principle.

Consistency with basic wave-like behaviors is guaranteed by the rigorous quantum mechanical treatment of the incident particle in the models leading to (15). The manifestation of such behavior in (14) is, however, noteworthy since no wave-like properties of the particle were considered in the derivation of (14). In the context of the thermodynamic analysis in Section I there exists an alternative physical interpretation of the variation (14) with $\varepsilon/T_1$. The total change in entropy $\Delta S_a$ associated with particle absorption is given by $(\varepsilon/T_1)-s$. It follows from basic thermodynamic principles that the probability for absorption must decrease as $\Delta S_a$ decreases. If $T_1$ and $s$ are constant then $\Delta S_a$ always decreases with decreasing $\varepsilon$. The probability (14) for thermodynamic reflection must accordingly increase with decreasing $\varepsilon$. Note also that the change in entropy $Q/T$ associated with the absorption of a quantity $Q$ of energy by any system at constant temperature $T$ also decreases with decreasing $Q$. The variation of the thermodynamic reflection probability with $\varepsilon/T_1$, for a fixed $s$, is therefore a natural consequence of thermodynamic principles.

Although it is possible to interpret the relationship between (14) and $\varepsilon/T_1$ independently of any wave-like properties, it is nonetheless remarkable that the wave-like behavior associated inherently with quantum mechanics is the same behavior that may preserves the GSL in certain interactions. It is also suggestive that one of the earliest intimations of the wave-particle duality is found in Einstein's formula giving the probability for a fluctuation of radiative energy within a system [15],[16],[17]. Einstein obtained the fluctuation formula by considering the variation in the number $W$ of microscopic complexions in terms of variations in the entropy $S$ associated with a given fluctuation, where $W=\exp(S/k)$ [17]. The probabilities (13) and (14) were obtained from the same basic principles. It is plausible that the manifestation of wave-like behavior in the probability for thermodynamic reflection represents the same physical unity implied by the wave-like behavior in Einstein's fluctuation formula.

It is appropriate to address an additional point concerning entropy and information. The thermodynamic entropy of a system is proportional to the number of bits of information registered by system [4]. A bit of information has two possible microstates, and the entropy $s_0$ associated with one bit is consequently $k_B\ln(2)$ [4]. In the foregoing analysis entropy was treated as a continuous variable. There may exist situations in which the quantization of information is important. For instance, in the context of Section I, if $0<|\Delta S_a|<s_0$ then the behavior of a system like $X$ may deviate significantly from models in which entropy is treated continuously. The quantization of information may also imply a certain limit for the probability for thermodynamic reflection. If any given particle must register at least one bit of information then $s$ may be no less than $s_0$. In the limit as $\varepsilon$ vanishes and $s$ approaches $s_0$, the probability function (14) approaches 2/3.

Acknowledgments: This work benefited from discussions with J. Katz and anonymous referees.
___________________

**References:**
[1] J. Bekenstein, Contemp. Phys. **45**:31-43 (2003)
[2] J.M. Bardeen, B. Carter, S.W. Hawking, "The four laws of black hole mechanics". Comm. Math. Physics **31**(2) 161-170 (1973)
[3] P.C.W. Davies, T.M.Davis, "How Far Can the Generalized Second Law Be Generalized?" Foundations of Physics, 32(12) 1877-1889 (2002)
[4] S. Lloyd, "Computational Capacity of the Universe" Phys. Rev. Lett. 88(23)1-4 (2002)
[5] R. Brustein, "The Generalized Second Law of Thermodynamics in Cosmology," Phys.Rev.Lett. **84** 2072-2085 (2000)
[6] J.D. Bekenstein, "Black holes and entropy". Phys. Rev. D **7** (8): 2333–2346 (1973).
[7] J.B. Hartle and S.W. Hawking, Path integral derivation of black hole radiance, Phys. Rev. D 13, 2188
(1976).
[8] S. Mendoza et al., Entropy, 11, 17-31 (2009)
[9] M. Yu. Kuchiev, arXiv:gr-qc/0310008.
[10] M. Yu. Kuchiev, Europhys. Lett. **65**, 445 (2004).
[11] M. Yu. Kuchiev, Phys. Rev. D **69**, 124031 (2004).
[12] M. Yu. Kuchiev and V. V. Flambaum, Phys.Rev. D **70** 044022 (2004)
[13] M. Yu. Kuchiev and V. V. Flambaum, arXiv:gr-qc/0407077.
[14] V.V. Flambaum, arXiv:gr-qc/0408013
[15] A. Kojevnikov, "Einstein's fluctuation formula and the wave-particle duality,"
In: *Einstein studies in Russia*. Yuri Balashov, Vladimir Vizgin (eds.). Einstein Studies, Vol. 10. Boston: Birkhäuser, ISBN 0-8176-4263-3, 2002, p. 181 – 228
[16] Sandor Varro, "Einstein's fluctuation formula. A historical overview," Fluctuation and Noise Letters, **6** No3 R11-R46 (2006)
[17]A. Einstein, "Zum gegenwärtigen Stand des Strahlungsproblems", Phys. Zeitschr. **10** pp185-193 (1909)